# Effect of impurities on morphology and growth mode of (111) and (001) epitaxial-like ScN films

Arnaud le Febvrier,* Nina Tureson, Nina Stilkerich, Grzegorz Greczynski, Per Eklund

Thin Film Physics Division, Department of Physics, Chemistry, and Biology (IFM), Linköping University, SE-581 83 Linköping, Sweden

* corresponding author: arnaud.le.febvrier@liu.se

# Abstract

ScN material is an emerging semiconductor with an indirect bandgap. It has attracted attention for its thermoelectric properties, use as seed layers, and for alloys for piezoelectric application. ScN or other transition metal nitride semiconductors used for their interesting electrical properties are sensitive to contaminants, such as oxygen or fluorine. In this present article, the influence of depositions conditions on the amount of oxygen contaminants incorporated in ScN films were investigated and their effects on the electrical properties (electrical resistivity and Seebeck coefficient) were studied. The epitaxial-like films of thickness 125 ± 5 nm to 155 ±5 nm were deposited by D.C.-magnetron sputtering on c-plane $Al_2O_3$, MgO(111) and r-plane $Al_2O_3$ at a substrate temperature ranging from 700 °C to 950 °C. The amount of oxygen contaminants presents in the film, dissolved into ScN or as an oxide, was related to the adatom mobility during growth, which is affected by the deposition temperature and the presence of twin domain growth. The lowest values of electrical resistivity of 50 µΩ cm were obtained on ScN(111)/MgO(111) and on ScN(001)/r-plane $Al_2O_3$ grown at 950°C with no twin domains and the lowest amount of oxygen contaminant. At the best, the films exhibited an electrical resistivity of 50 µΩ cm with Seebeck coefficient values maintained at -40 µV $K^{-1}$, thus a power factor estimated at 3.2 ×$10^{-3}$ W $m^{-1}$ $K^{-2}$ (at room temperature).

# Introduction

Transition metal nitride thin film are widely investigated for their mechanical [1-4], plasmonic [5], piezoelectric properties [6-8], or used as diffusion barriers [9, 10] and more recently for their thermoelectric properties [11]. For Physical vapor deposition (PVD) such as sputtering or arc deposition, it is important to control the purity of the final film. With such techniques, high vacuum or ultra-high vacuum chambers with base pressure of $10^{-6}$ to $10^{-8}$ Pa are used to minimize the contaminations during the deposition process. Different aspects of the deposition process can be controlled in order to minimize contamination, such as the cleaning processes of chamber and substrates, bake-out, base pressure and purity of target / gas. Nevertheless, even at the best conditions of cleanliness, some transition metals which are more sensitive than others to contamination (oxygen, carbon) will lead to incorporation of impurities in the film during deposition.

Among the transition metals, scandium has one of the highest affinities for oxygen [12]. Presence of oxygen in the final film occurs most likely during the deposition process [12], rather than as a result of oxidation of the film by grain boundary diffusion during postgrowth exposure to air[13]. Another source of contaminants for ScN is the target. Due the production process, scandium target contains fluorine impurity which is typically found in the ScN film [14-18]. Oxygen impurities incorporated into ScN influence the electronic properties of ScN films [12, 14, 16, 19-24]. Scandium nitride as many of the transition metal nitride materials has a rock-salt structure (B1) and exhibit high hardness and high melting point [25, 26]. ScN is a degenerated semiconductor with an indirect bandgap of 0.9-1.6 eV [12, 22] and shows a n type behavior with carrier concentration varying from $10^{18}$ to $10^{20}$ cm$^{-3}$ [22, 27]. The presence of defects, impurities or variation of composition into ScN may affect drastically the electronic and thermoelectric properties of the final film.

For thermoelectrics, control of contaminants is essential for control of the electrical properties of the film. The efficiency of a thermoelectric material is defined by its dimensionless *figure of merit*, $ZT = \frac{S^2 \cdot \sigma}{\kappa} \cdot T$, where S is the Seebeck coefficient, σ is the electrical conductivity, κ is the thermal conductivity and T is the absolute temperature [28, 29]. The higher the thermoelectric *figure of merit*, i.e., the higher the power factor ($S^2 \cdot \sigma$) and/or the lower the thermal conductivity, the more efficient the energy conversion. The origin of the large variation, thermoelectric properties obtained on ScN can be explained by a modification of the density of states (DOS) of ScN around the Fermi level with respect to the presence of points defects or impurities [30]. Kerdsongpanya et. al. have shown theoretically the possibility to alter the thermoelectric properties by the presence of contaminants [30]. Incorporation of oxygen ( ≤ 2 at.%) into the ScN cell causes a shift of the Fermi level into the conduction band without affecting the DOS enhancing the carrier concentration and electrical conductivity [11, 14, 20, 21, 30]. When the amount of contaminants increases, these electrical properties may deteriorate, and at even higher level, secondary phases are formed such as $Sc_2O_{3-x}$ [12, 14, 16, 20, 31, 32]. Our previous study on wet-cleaning of MgO(001) substrates have shown the important influence of the substrate cleaning process used prior to deposition on the reduce of scandium oxide formation into the film [31]. Generally, for ScN deposited by reactive sputtering, the lower the base pressure of the chamber, the lower is the content of oxygen incorporated into ScN [12, 20, 21]. Several papers [12, 14, 16, 20, 21, 25, 30-33] reported oxygen contaminations from below one to several percent into ScN but there is still a need to elucidate the effect of oxygen incorporation on the growth mechanism and the effect of the orientation of the film and the presence or not of twin domains during growth. Understanding of the oxygen incorporation into ScN film is a key point for control and improvement of its electrical and thermoelectric properties.

In this study, ScN thin film were studied to improve the understanding of the film growth process and the control of incorporation of oxygen. Different conditions of temperature (700 to 950 °C) and different substrates (c-plane Al$_2$O$_3$, MgO(111) and r-plane Al$_2$O$_3$) were used to deposit ScN film in a high vacuum chamber. Influence of the deposition conditions as well as the substrates used influences the incorporation of oxygen into ScN cell and/or oxide. The substrates used, for this study, was chosen for their dielectric character allowing further electrical and thermoelectric characterization and those ones are also commonly used to deposit thermoelectric thin films. Despite their nature, different orientations were considered in order to grow the film along different orientations and deposit the films under different growth mode. The effect of the temperature used, surface terminated atoms, twin domain growth, adatom mobility during deposition and oxygen content on the electrical and thermoelectric properties of ScN is investigated.

## Materials and methods

ScN thin films were deposited using D.C. reactive magnetron sputtering in a high vacuum chamber ($10^{-7}$ Pa) using a 2-inch Sc target (MaTek: Sc 99.5%) in an Ar/$N_2$ (flow ratio 75% Ar / 25% $N_2$) sputtering-gas mixture. The pressure during depositions was kept at 0.27 Pa (2 mTorr) as well as the power at 125 W. The chamber is described elsewhere [34]. 10 mm x 10 mm one side-polished substrates of c-plane $Al_2O_3$, MgO(111) and r-plane $Al_2O_3$ (Alineason Materials & Technology) were used and simultaneously inserted in the chamber, on the same holder for deposition. The films were deposited during 60 min under constant rotation and at three different temperatures of deposition ($T_D$): 700, 820 and 950 °C. Prior to deposition the substrates were cleaned using detergent steps, then 10 min with acetone in ultrasonic bath and then repeated with ethanol and blown dry with a $N_2$-gun. The detergent steps are described elsewhere [31].

Structural characterization was performed by X-ray diffraction (XRD) on a PANalytical X'Pert PRO in Bragg Brentano configuration (with continuous rotation of the sample during the measurement) equipped with a Cu $K_\alpha$ radiation source and a four-circle texture instrument (PANalytical X'Pert MRD) equipped with a parallel beam Cu $K_{\alpha 1}$ radiation in $\theta-2\theta$, $\omega$ and $\varphi$ scan modes and for the x-ray reflectivity measurement. The morphology and the cross sections of the film were observed by a scanning electron microscope (SEM, LEO Gemini 1550, Zeiss).

XPS spectra were obtained using an Axis Ultra DLD instrument from Kratos Analytical (UK) with the base pressure during spectra acquisition of $1.1 \times 10^{-9}$ Torr ($1.5 \times 10^{-7}$ Pa), with a monochromatic Al K$\alpha$ radiation (h$\nu$ = 1486.6 eV). The anode power was set to 150 W. Prior to analyses all samples were sputter-cleaned with 0.5 keV Ar+ ion beam incident at the 20° angle from the surface and rastered over the area of $3 \times 3$ mm$^2$. All spectra were collected from the area of $0.3 \times 0.7$ mm$^2$ and at normal emission angle using a low-energy electron gun to compensate for the sample charging. The analyzer pass energy was set to 20 eV which results in the full width at half maximum of 0.55 eV for the Ag $3d_{5/2}$ peak. XPS data were treated using KolXPD fitting software [35]. To avoid doubts related to using the C 1s peak of adventitious carbon as a charge reference [36, 37]. all spectra were aligned to the N 1s peak of Sc-N set at 396.5 eV. The latter procedure results in N 1s and Sc 2p binding energy values which are consistent with the NIST data base [38]. Peak fitting was performed for all spectra using Voigt functions and a Shirley background. Details of the peak fitting procedure are described in the supplementary information.

The electrical resistivity of the sample was determined indirectly by measuring the sheet resistance of the film with a four-point probe Jandel RM3000 station which was multiplied by the thickness of the films obtained from the cross-sectional SEM imaging. The Seebeck voltage was measured on a homemade Seebeck voltage measurement setup and performed at atmospheric pressure. The samples were electrically isolated on which two copper electrodes separated by 8 mm were connected to the surface of the sample (film) and connected to a multimeter with a resolution of 0.01 mV. A temperature differential of 47 °C was applied over the film using a heated metal tip on one electrode and the other one kept at room temperature ($T_{hot}$ = 74 °C and $T_{cold}$ = 27 °C). The temperature gradient and the Seebeck voltage were measured after temperature stabilization (holding time: 10 min).

## Results

Fig. 1 (a, b, c) shows Bragg-Brentano θ-2θ scans of the different ScN films deposited at different temperatures (700, 820 and 950 °C) on c-plane $Al_2O_3$, MgO(111) and r-plane $Al_2O_3$. For all thin films, the most intense diffraction peak detected from the film is the 111 reflection of cubic ScN rock salt structure. The differences between the nine films are noticeable by the orientations and/or presence of a secondary phase. The films deposited at 700 °C on c-plane $Al_2O_3$ and MgO(111) substrates are preferentially (111) oriented, along with the presence (001) and (110) oriented ScN grains. On the same films, $Sc_2O_3$ phase was detected with a preferential (332) orientation. The film deposited on the r-plane $Al_2O_3$ was preferentially (001) oriented and no $Sc_2O_3$ peak was detected on the film deposited at 700 °C. At a higher $T_D$ of 820 °C, no peak of the $Sc_2O_3$ was detected, and the films were composed of ScN grains with their respective preferential orientations ((111) or (001)) and the presence of (001) or (110) secondary orientations of ScN. At a $T_D$ of 950°C, only one diffraction peak was detected from the thin film: films deposited on c-plane $Al_2O_3$ and MgO(111) substrates were (111) oriented and the film on r-plane $Al_2O_3$ substrate was (001) oriented. Fig. 1(d, e, f) shows φ-scans measured on the $\{220\}_{ScN}$ reflection at azimuth angle ψ=35.2° ((111) oriented film) and ψ=45° ((001) oriented film). On c-plane $Al_2O_3$ substrate, six peaks are present and separated by 60° due to twin domains of an (111) oriented cubic symmetry crystal film grown on a hexagonal crystal structure. On MgO(111) substrate, three peaks are present and separated by 120° characteristic of a (111) cubic symmetry film. On r-plane $Al_2O_3$ substrate, four peaks are visible and separated by 90 °C characteristic of a (001) oriented cubic symmetry film. An increase of $T_D$ increases the intensity of diffraction in the φ-scans and the diffraction peaks become narrower.

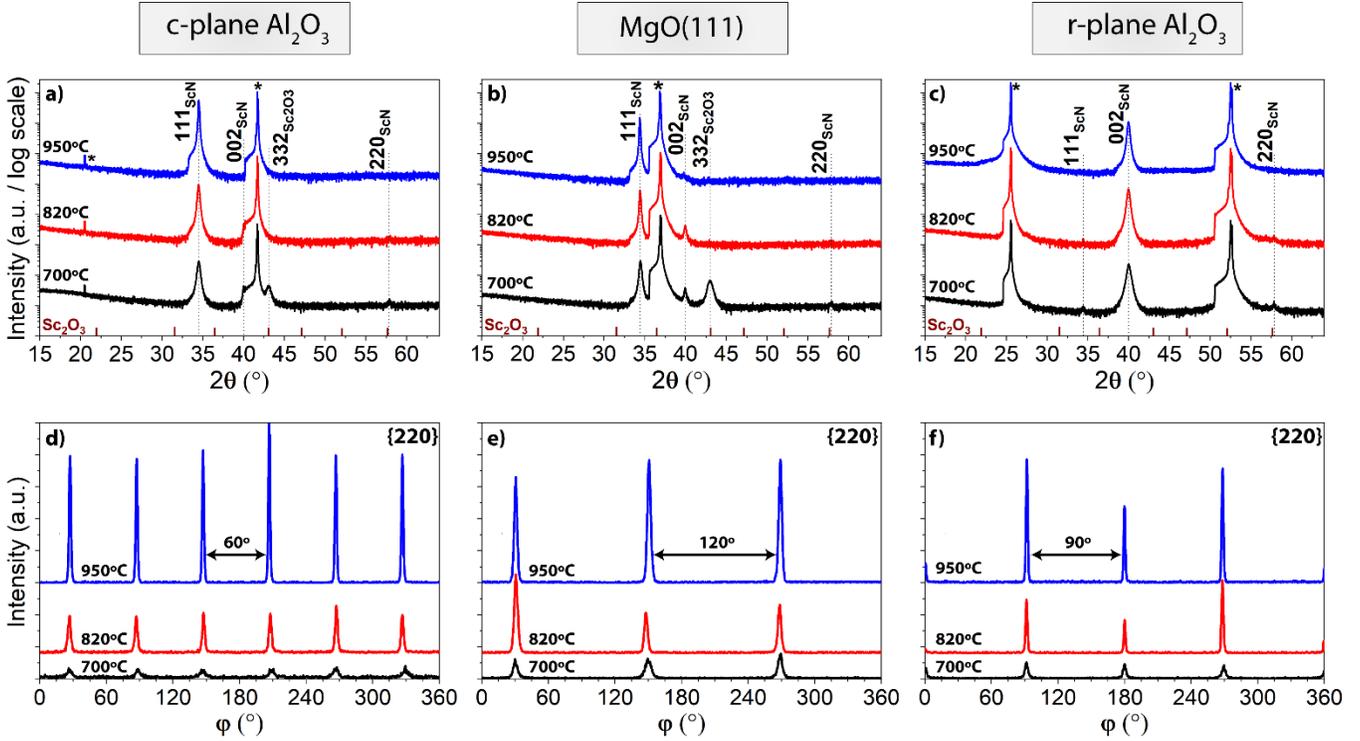

**Figure 1:** X-ray diffraction patterns (θ-2θ scan) and the corresponding φ-scans of the {220} reflections of the oriented ScN film deposited at different temperatures (700 °C, 820 °C and 950 °C) on: (a, d) c-plane $Al_2O_3$, (b, e) MgO(111) and (c, f) r-plane $Al_2O_3$. Peaks marked as ∗ correspond to the respective substrate reflection peaks: 0003 ($Al_2O_3$), 0006 ($Al_2O_3$), 111 (MgO), $10\bar{1}2$ ($Al_2O_3$) and $20\bar{2}4$ ($Al_2O_3$). of the $Sc_2O_3$ (ICDD data: 00-005-0629) are represent by small ticks.

The detailed results from the XRD analysis and the thickness values are listed in table 1. The thickness is reduced from 155 ± 5 nm at a $T_D$ of 700 °C to 125 ± 5 nm at a $T_D$ of 950 °C. The mass density deduced from the X-ray reflectivity measurements varies from 3.75(1) g cm$^{-3}$ at low temperature to 4.09(1) g cm$^{-3}$ at high temperature. The densities obtained can be compared to the ScN bulk values of 4.25(2) [39]. The film deposited on c-plane $Al_2O_3$ do not exhibit a clear difference of the cell parameter when the $T_D$ increases from 700 °C to 950 °C with a value close to 4.504(4) Å. The ScN cell parameter deposited on MgO(111) increases from 4.503(4) Å at $T_D$ of 700 °C to 4.512(4) Å at 850 °C and stay constant for higher temperature. Similar observation can be made for the film deposited on r-plane $Al_2O_3$. The crystal quality can be evaluated from the measurement of the Δω and Δφ on the ω-scans and φ-scans of the preferential orientation. The lower the values of the Δω and Δφ, the higher the crystalline quality for a certain orientation. All films exhibited diffraction peaks on the φ-scans demonstrating a degree of ordering of the grains in the plane of the substrate. The term "epitaxial-like" can be used to describe this, meaning that all grains are epitaxially related to the substrate, and domain-growth with two different stacking sequences occurs, but there is no global epitaxy. The crystal quality differs between substrates, with a higher crystal quality on c-plane $Al_2O_3$ and r-plane $Al_2O_3$ substrates than on MgO(111) substrate. Nevertheless, all films are epitaxially-like grown with higher quality at higher $T_D$ observed by a reduction of the Δω and Δφ.

**Table 1:** Details from XRD analysis of ScN films deposited at different temperatures (700°C, 820°C and 950°C) on: c-plane $Al_2O_3$, MgO(111) and r-plane $Al_2O_3$. preferential orientations of the film; cell parameter calculated from the main peak; associated ω-scan and φ-scan FWHM values of the preferential orientation and thickness are reported.

| Deposition temperature ($T_D$) | Substrate | ScN preferential orientation | $a_{ScN}$ (± 0.004 Å) | ω-scan FWHM (Δω in °) | φ-scan FWHM (Δφ in °) | XRR density (g cm$^{-3}$) (± 0.01) | Thickness (± 5 nm) |
|---|---|---|---|---|---|---|---|
| 950 °C | c-plane $Al_2O_3$ | (111) | 4.503 | 0.4 | 1.6 | 4.09 | 120 |
|  | MgO(111) | (111) | 4.512 | 1.5 | 2.9 | 4.08 | 130 |
|  | r-plane $Al_2O_3$ | (001) | 4.509 | 1.4 | 1.7 | 4.09 | 125 |
| 820 °C | c-plane $Al_2O_3$ | (111) | 4.504 | 0.8 | 2.6 | 3.91 | 130 |
|  | MgO(111) | (111) | 4.512 | 1.6 | 3.3 | 4.01 | 120 |
|  | r-plane $Al_2O_3$ | (001) | 4.509 | 1.3 | 1.7 | 4.01 | 140 |
| 700 °C | c-plane $Al_2O_3$ | (111) | 4.504 | 1.0 | 3.6 | 3.75 | 160 |
|  | MgO(111) | (111) | 4.503 | 2.1 | 5.1 | 3.85 | 150 |
|  | r-plane $Al_2O_3$ | (001) | 4.505 | 2.1 | 3.4 | 3.81 | 155 |

Fig. 2 shows SEM images of the film surface morphologies. The surface morphology of the film grown on the three substrates at 950 °C is dense and smooth. The films on c-plane $Al_2O_3$ and MgO(111) are composed of small pyramidal shaped grains of 10-15 nm characteristic of a (111) oriented cubic symmetry material. On c-plane Al$_2$O$_3$, the dark areas between grains observed at the surface of the film are due to the twin domain growth of the film as opposed to the one on MgO(111) substrate. The film deposited on r-plane Al$_2$O$_3$ is composed of regular well-organized square-shaped grains with a size of 35-40 nm characteristic of a (001) oriented cubic symmetry material. The morphology of the films deposited at 750 °C present a rough surface with large grain size. For all three substrates, the crystallites present a certain disorder at the surface of the film which shows a trend of texturing along the [111] or [001] directions. A film deposited at a T$_D$ of 820 °C can be described as an intermediate situation.

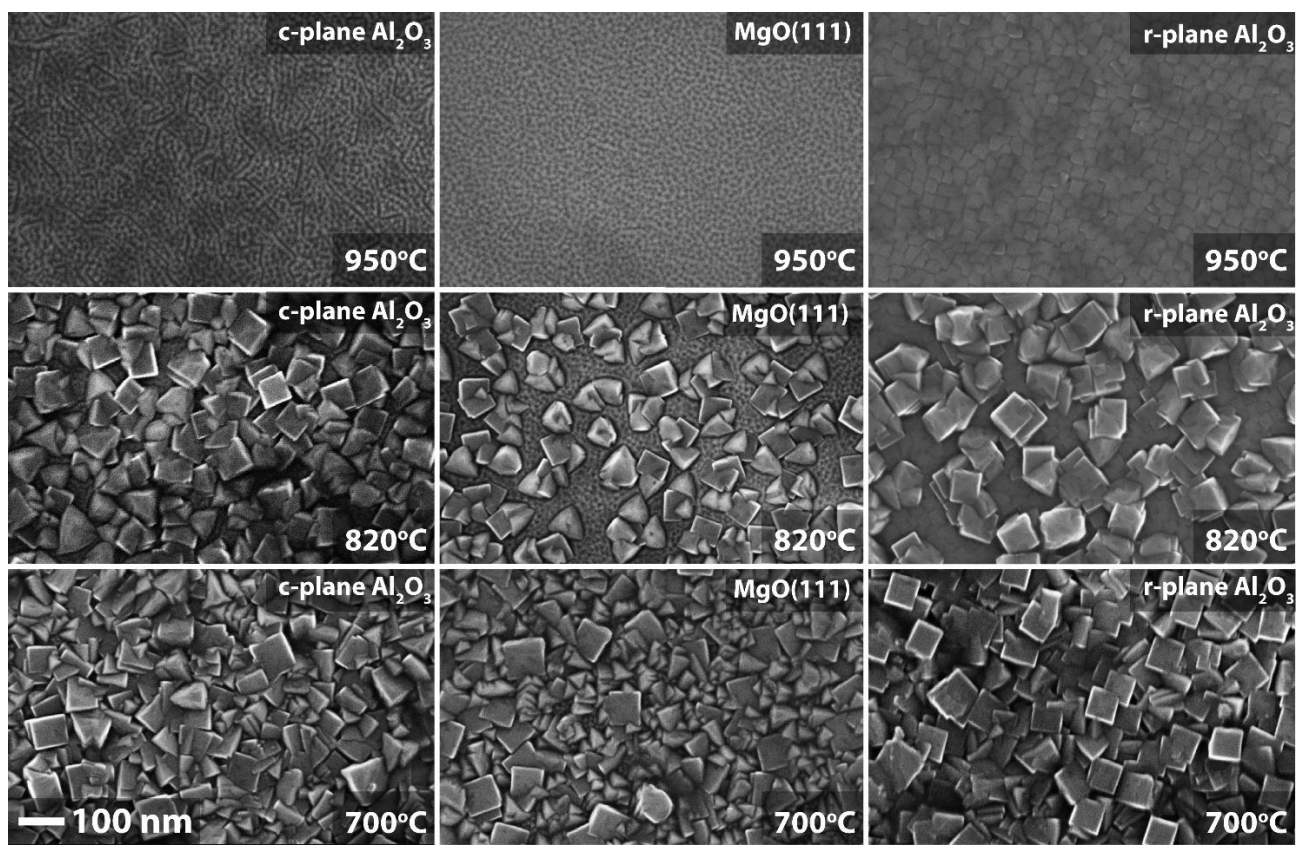

**Figure 2:** top view SEM image / surface morphology of the ScN films deposited at different temperatures (700 °C, 820 °C and 950 °C) on: c-plane Al$_2$O$_3$, MgO(111) and r-plane Al$_2$O$_3$.

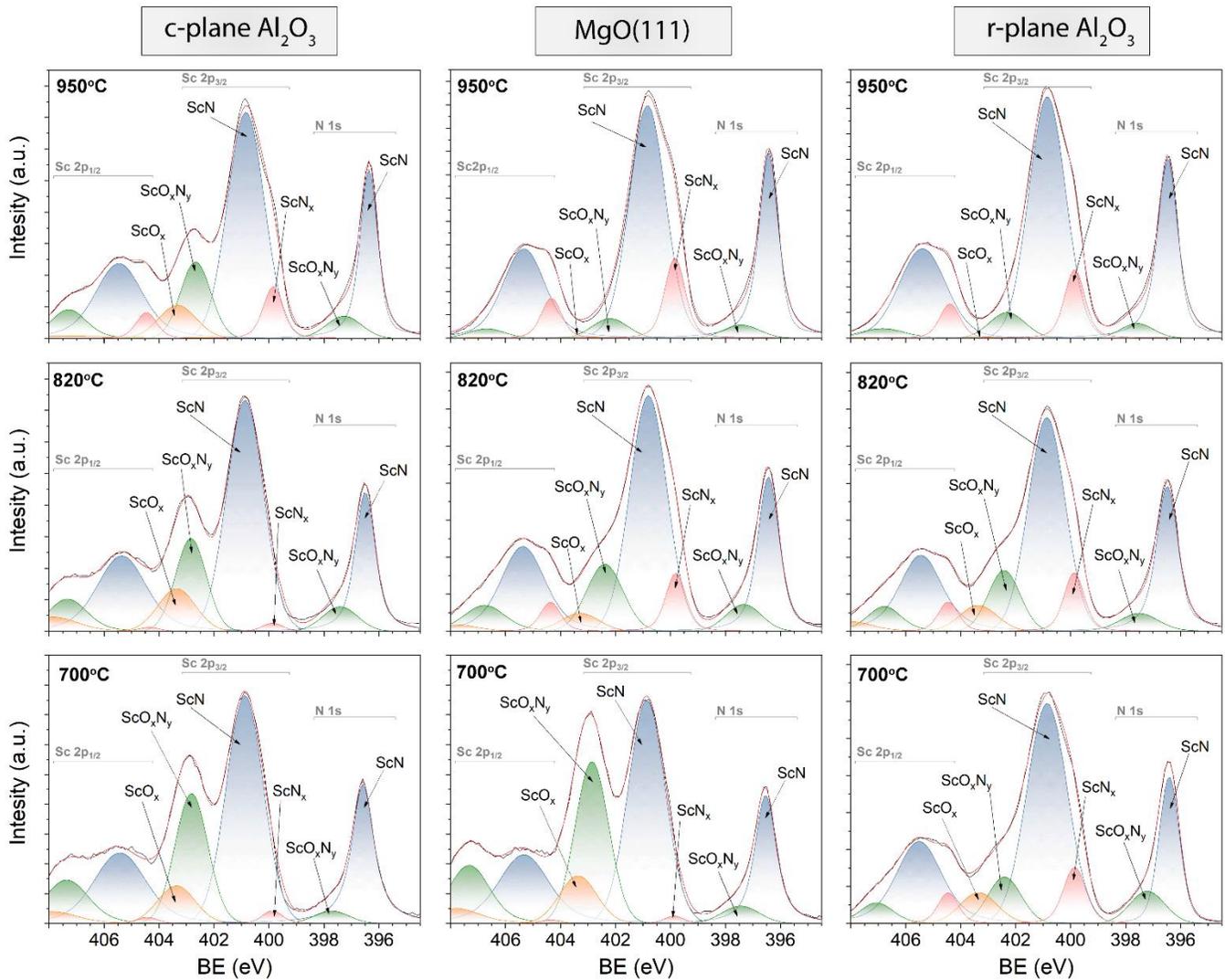

**Figure 3:** XPS peak models of Sc 2p and N 1s core level spectra acquired from films deposited on c-plane Al$_2$O$_3$, MgO(111), and r-plane Al$_2$O$_3$ at 700 °C, 820 °C and 950 °C.

Fig. 3 displays the detailed XPS spectra of the Sc 2p and N 1s core levels together with the deconvoluted peaks. The N 1s peak is composed for all films of a main N-Sc peak at 396.5 eV and a shoulder peak on the high-BE side (397.4 eV) corresponding to nitrogen in an oxynitride (N-(ScO$_x$N$_{1-x}$)). The Sc 2p core level peak has several contributions: ScN peak characteristic of the ScN material situated at 400.8 eV [38]; a Sc-N peak of substoichiometric ScN$_x$ (x < 1) at 399.8 eV formed as a result of preferential sputtering of lighter N atoms [40, 41]. The oxygen present in the film either as ScO$_x$ or dissolved in ScN (ScO$_x$N$_{1-x}$) gives rise to two more peaks in the Sc 2p spectra at 403.3 and 402.2-402.9 eV, respectively. With increasing growth temperature, the ScN$_x$ peak becomes more intense as the effective area cleaned by the Ar$^+$ beam increased due to the reduced surface roughness. Thehis increasing of temperature is also responsible for a decreasing intensity of the Sc 2p components of the oxide and oxynitride.

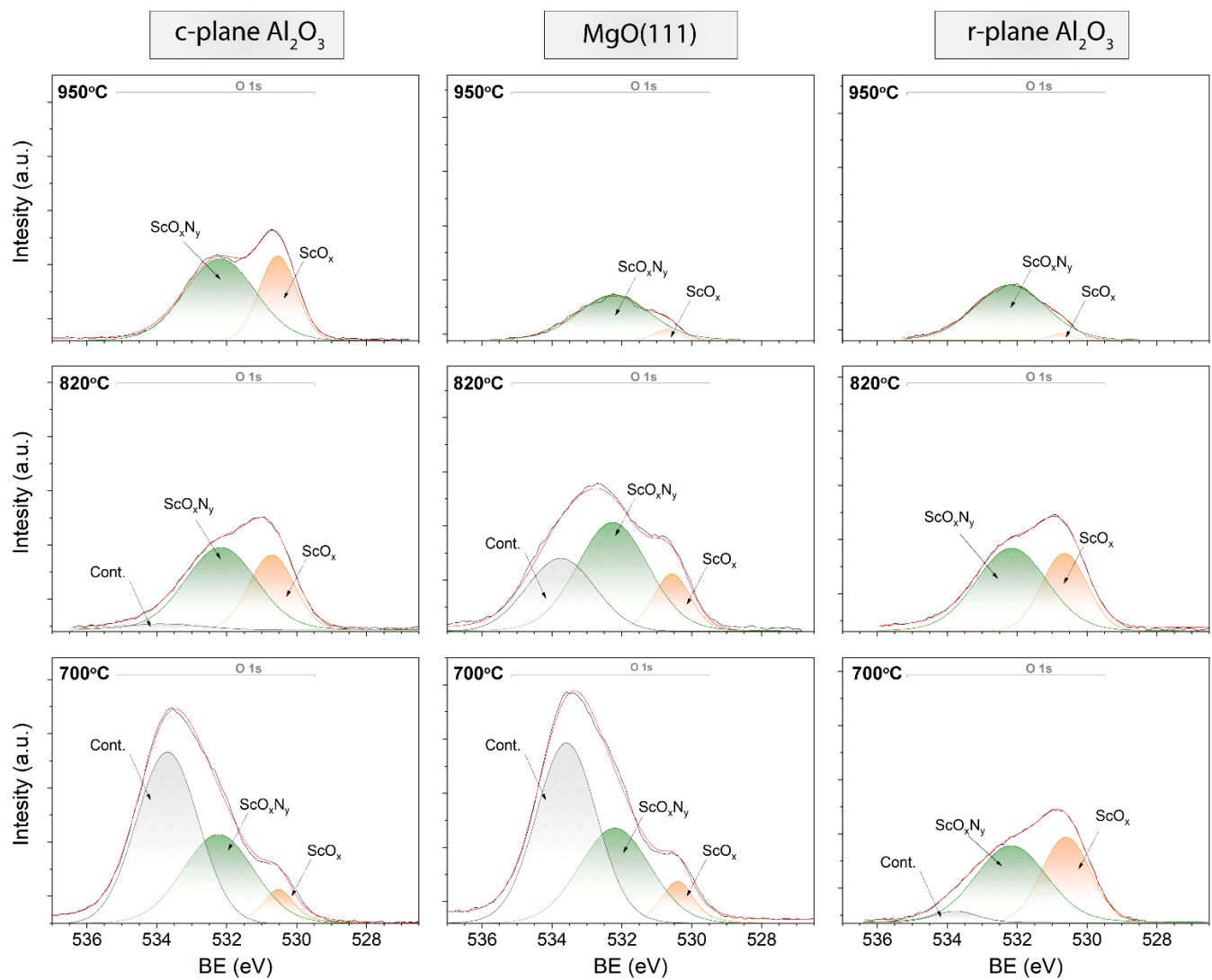

**Figure 4:** XPS peak models of O 1s core level spectra acquired from films deposited on c-plane Al$_2$O$_3$, MgO(111), and r-plane Al$_2$O$_3$ at 700 °C, 820 °C and 950 °C.

Fig. 4 displays the detailed O 1s XPS spectra together with the peak models. Clearly, numerous contributions indicate that oxygen is present in several chemical states. A peak at 530.4 eV is assigned to the oxide $ScO_x$), followed by a higher BE component at 532.2 eV due to oxynitride $ScO_xN_{1-x}$. Finally, the peak at the highest BE (533.8 eV) is due to the contaminating species including CO, $CO_2$, and $H_2O$ which remain at the surface due to incomplete cleaning of rougher films, as described above. The later component disappears in case of the 950 °C sample, which has the lowest surface roughness, and, hence, can be properly cleaned with the $Ar^+$ ion beam.

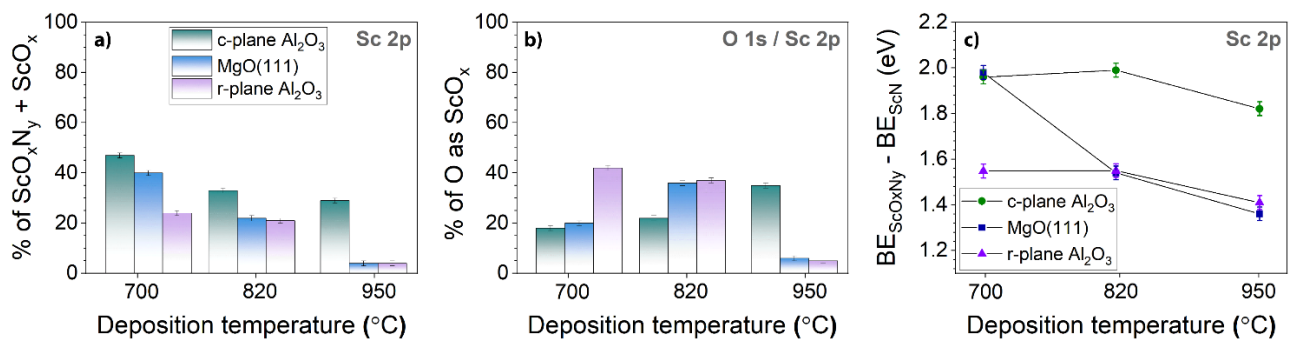

**Fig. 5:** Summary of the Sc 2p XPS peak modelling performed for films grown on c-plane $Al_2O_3$, MgO(111) and r-plane $Al_2O_3$ and at 700 °C, 820 °C and 950 °C. a) fraction of the scandium in $ScO_xN_y$ + $ScO_x$; b) fraction of oxygen detected as an oxide $ScO_x$; c) binding energy (BE) difference between the ScN and the $ScO_xN_y$ peaks in the Sc 2p spectra.

Figure 5 shows the summary of the information extracted from the XPS analysis performed on the nine ScN films (see Fig. 3 and 4). Fig. 5a displays the total fraction of the scandium in an oxynitride $ScO_xN_{1-x}$ (x > 0) plus oxide $ScO_x$ and Fig. 5b represents the $ScO_x$ fraction detected in the Sc 2p or O 1s peak. The films on r-plane $Al_2O_3$ and on MgO(111) grown at 950 °C contain the lowest amount of Sc bonded to oxygen (only 4% of the total area under the Sc 2p spectra), of which 95% corresponds to an oxynitride $ScO_xN_{1-x}$ (x > 0). For the film on c-plane $Al_2O_3$ grown at 950°C, 30 % of Sc atoms is bonded to oxygen, of which 60% are in the oxynitride $ScO_xN_{1-x}$ (x > 0) and 40% in $ScO_x$. At lower $T_D$ the amount of scandium bonded to oxygen increases for all three substrates used. At a $T_D$ = 700°C, 24% of Sc is bonded to oxygen in films on r-plane $Al_2O_3$, while up to 40-47% of Sc-O bonds are detected for the films on c-plane $Al_2O_3$ and MgO(111) substrates. Fig. 5c, shows the evolution of the relative BE difference between the oxynitride and the nitride components in the Sc 2p core level spectra. The total amount of oxygen and the relative position of the oxynitride peak ($ScO_xN_{1-x}$ (x > 0)) in Sc 2p core level peak vary with the temperature of deposition. The higher the oxygen content (The lower the temperature), the higher the shift of the oxynitride peak towards higher BE, which indicates that the O-to-N ratio in oxynitride increases.

**Table 2:** Thin film compositions deduced from XPS analyses.

| Deposition temperature ($T_D$) | Substrate | Elemental composition (at.%) (± 0.5%) | | | | | | | Ratio N/Sc | Ratio (N+O+F)/Sc |
|---|---|---|---|---|---|---|---|---|---|---|
| | | C | F | O Cont. | O $ScO_x$ | O $ScO_xN_{1-x}$ | N | Sc | | |
| 950 °C | c-plane $Al_2O_3$ | 2.3 | 0.7 | 1.2 | **5.6** | **10.3** | 34.4 | 45.5 | **0.75** | 1.12 |
| | MgO(111) | 0.4 | 1.0 | 0.0 | **0.3** | **4.4** | 45.3 | 48.5 | **0.93** | 1.05 |
| | r-plane $Al_2O_3$ | 0.2 | 0.9 | 0.0 | **0.2** | **3.6** | 46.7 | 48.4 | **0.97** | 1.06 |
| 820 °C | c-plane $Al_2O_3$ | 14.9 | 0.8 | 8.7 | **6.4** | **11.4** | 23.5 | 34.4 | **0.68** | 1.22 |
| | MgO(111) | 7.1 | 1.2 | 0.8 | **2.8** | **9.9** | 33.9 | 44.3 | **0.76** | 1.08 |
| | r-plane $Al_2O_3$ | 4.5 | 1.3 | 0.4 | **4.3** | **8.1** | 35.9 | 45.5 | **0.79** | 1.09 |
| 700 °C | c-plane $Al_2O_3$ | 21.6 | 0.7 | 21.7 | **3.1** | **14.0** | 13.5 | 25.4 | **0.53** | 1.23 |
| | MgO(111) | 24.9 | 0.4 | 21.2 | **4.0** | **12.1** | 14.4 | 23.0 | **0.63** | 1.34 |
| | r-plane $Al_2O_3$ | 3.8 | 1.1 | 0.8 | **5.3** | **7.30** | 36.1 | 45.6 | **0.79** | 1.09 |

The compositions of the ScN film as estimated from XPS, are listed in the table 2. Fluorine impurities were detected at around 1% or less. The anion/cation ratio varies from 1.06 for the films with the lowest content of oxygen to 1.23-1.34 for the films with the highest amount of oxygen. A higher $T_D$ results in thin film containing a low amount of oxygen (4-5%) with a high N/Sc ratio between 0.93 and 0.97 for the film on MgO(111) and r-plane $Al_2O_3$, respectively. The oxygen detected in those two films is mostly present as oxynitride. The film deposited on c-plane $Al_2O_3$ had a lower N/Sc ratio of 0.75 and contained a significant amount of oxygen, 16 %, of which 10.3 % is due to $ScO_xN_{1-x}$ (x > 0) and 5.6% due to an oxide. A lower $T_D$ yields ScN films containing higher content of oxygen (10-15 %), found as oxynitride and oxide and associated with lower ratio of N/Sc between 0.53 (c-plane $Al_2O_3$ substrate) and 0.79 (r-plane $Al_2O_3$ substrate). The XPS analysis of the surface of the films grown at 700 °C, and the one on c-plane $Al_2O_3$ grown at 820°C revealed a high content of carbon even after a sputter cleaning process which was kept identical for all samples. This amount of carbon is an indication of the effectiveness of the sputter cleaning, which is more difficult to achieved due to the high roughness and the surface morphology of the film, leading to shadowing on the significant fractions of the surface. Note here that the films presenting a high content of carbon on surface are also the ones with a contamination peak contribution in the O 1s core level spectra (see Table 2 and Fig. 4, plus Fig. S2 in the supplementary information representing the C 1s core level spectra).

Fig. 6a shows the electrical resistivity of the thin films measured by four-point probe at room temperature. Differences between the substrates used are noticeable. The lowest electrical resistivities (45 to 370 μΩ cm) were measured on films deposited on MgO(111), then higher electrical resistivities (55 to 625 μΩ cm) were measured on films grown on r-plane $Al_2O_3$. Finally, on c-plane $Al_2O_3$, higher resistivities are observed (180 to 2700 μΩ cm). On the three substrates, all films exhibited a decrease of the electrical resistivity when the $T_D$ increases from 700 °C to 950 °C or when the oxygen content into the film decreases. Fig. 6b shows the Seebeck coefficient of the nine films. The films deposited on the three at the low $T_D$ exhibited similar Seebeck coefficients around -27 μV $K^{-1}$. At a $T_D$ = 950 °C, all films exhibited higher absolute values of Seebeck coefficient around -40 μV $K^{-1}$.

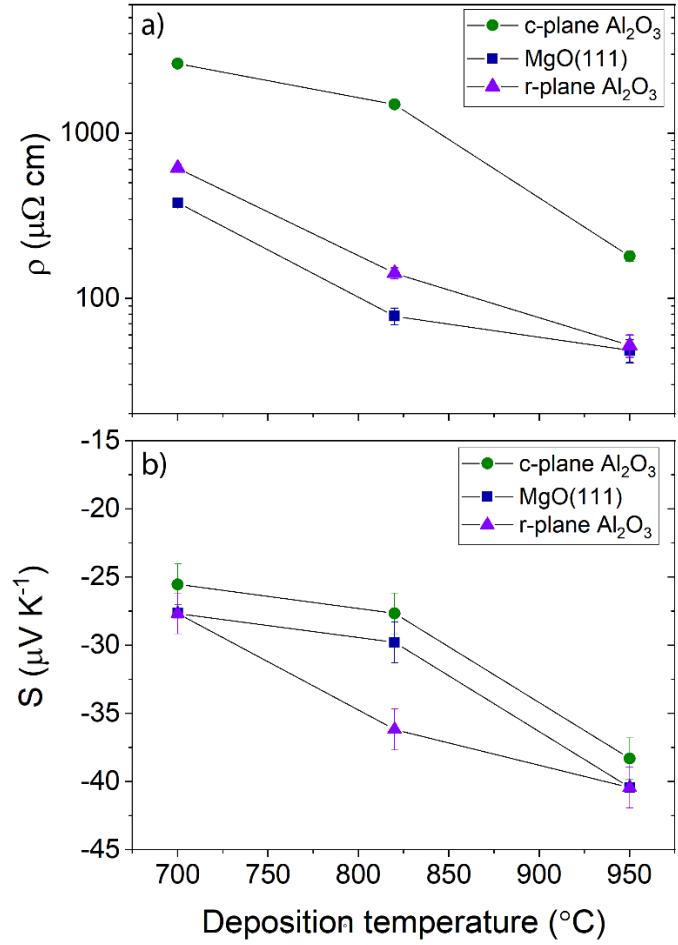

**Figure 6:** a) Room temperature electrical resistivity versus the temperature of growth ($T_D$) of the ScN films deposited on: c-plane $Al_2O_3$, MgO(111) and r-plane $Al_2O_3$. b) Estimated Seebeck coefficient at 50 °C versus the temperature of growth ($T_D$) of the same films. Measurements performed at an average temperature of 50 °C with a $\Delta T$ = 47 °C and at atmospheric pressure.

## Discussion

The different XRD analyses confirmed the rock salt crystal structure with a lattice constant close to the bulk values (4.5013 Å / ICDD 030656286). All ScN films were grown epitaxially on the three substrates with a (111) orientation on c-plane $Al_2O_3$ and MgO(111) and with a (001) orientation on r-plane $Al_2O_3$. The differences on cell parameter values between the nine films and the bulk value may originate from composition variation. An (111) orientation for the cubic cell ScN films is expected on c-plane $Al_2O_3$ with an epitaxial relationship of $[111]_{ScN}//[0001]_{Al2O3}$ and $[22\bar{1}]_{ScN}//[1000]_{Al2O3}$. The same orientation of the film is expected on MgO(111) with an epitaxial relationship of $[111]_{ScN}//[111]_{MgO}$ and $[10\bar{1}]_{ScN}//[10\bar{1}]_{MgO}$. The (001) orientation of ScN is observed r-plane $Al_2O_3$ with an epitaxial relationship of $[001]_{ScN}//[10\bar{1}2]_{Al2O3}$ and $[100]_{ScN}//[10\bar{1}1]_{Al2O3}$. Epitaxial growth will be possible with a compressive stress on c-plane $Al_2O_3$ (mismatch: +14%) and MgO(111) (mismatch: + 6%) and with a tensile/compressive stress for the film grown on r-plane $Al_2O_3$ (mismatch: -5%; +12%). Detailed scheme of the epitaxial relationship between the substrates and ScN are presented in supplementary information. The increase of temperature increased the epitaxial quality of the film with an increase of the main diffraction peak and a reduce of the Δω and Δφ associated to the main orientation of the film. At high temperature, the three films have an equivalent epitaxy quality (similar Δω and Δφ). The increase of the epitaxial quality is mainly due to the increase of energy brought during deposition by the temperature which increases the adatom mobility at the surface. As said before, the term *epitaxial-like* growth is used here to distinguish this from global epitaxy.

The morphology of the films at 950 °C varies depending of the substrates. This feature is highly correlated to the type of orientation and type of growth of the film. The morphology is also temperature-dependent on all three substrates with the suppression of mound grain growth when $T_D$ increases. The formation of this mound structure is commonly observed on ScN thin film and other transition metal nitrides [16, 20, 21, 25, 42-44]. Adatom mobility on a surface is affected by the presence of different defects or steps. The down-step motion of an adatom at a step edge is limited by the Ehrlich–Schwoebel barrier and favoring the uphill migration on terraces [45, 46]. Varying the barrier magnitude which depends of $T_D$, the kinetic surface roughness and faceting varies [47]. A high temperature of deposition provides sufficient adatom mobility to suppress the surface roughening. As observed in this study and on all three substrates, an increase of temperature decreases the number of exaggerated grain growth to dense and homogenous films composed of nanometric size grains at 950 °C.

Along with the improvement of the epitaxy quality when the $T_D$ increased from 700 °C to 950 °C, the nitrogen content and/or the ratio N/Sc increased from 0.75 to 0.97. The reactivity of scandium with nitrogen is enhanced at higher temperature as usually observed in D.C.-magnetron sputtering of nitride materials [48]. In the present case, a temperature of 950 °C leads to deposition a film with a ratio close to 1:1 (Sc:N) when grown on MgO(111) and r-plane $Al_2O_3$.

By XRD, the presence of oxide is only detected on the low temperature grown films on c-plane $Al_2O_3$ and on MgO(111). In opposite to the XRD with a Bragg-Brentano configuration where no oxide was detected, XPS revealed the presence of $ScO_x$ in the film grown on r-plane which has a higher amount of oxide detected than the other films grown at the same temperature. This can be due to:1) the low quantity of oxide ($\approx 5\%$), 2) a possibility that this oxide is amorphous or epitaxially oriented with a tilt regarding the normal of the surface and 3) the lower amount of oxide on the two other substrates was detected because of its preferential orientation (332) on c-plane sapphire and MgO (111). Note here that the measurement of XPS is a surface analysis technique. A larger total amount of oxide in the film or at the substrate/film interface was not excluded to explain the presence of the oxide diffraction peak for the film grown on c-plane $Al_2O_3$ and MgO(111) at 700 °C. Increasing the temperature of deposition lead to film of ScN containing lower amount of scandium oxynitride or oxide . The oxygen incorporation occurs mainly as dissolution into the ScN lattice and few percent as an oxide most likely at the grain boundary [20].

At a $T_D$ of 950 °C, even with the base pressure of $10^{-7}$ Pa, the oxygen incorporation occurred during deposition, mainly due to the residual water in the chamber leading to few at.% of oxygen in the film. At that highest $T_D$, less than 1 at.% of the oxygen detected is issued from an $ScO_x$ oxide and 3-5% of oxygen detected dissolved in ScN cell (as oxynitride) for the film grown on MgO(111) and r-plane $Al_2O_3$. Thus, the films deposited on MgO(111) and r-plane $Al_2O_3$, have similar impurities level and N/Sc ratio. The (111) oriented film epitaxially grown on c-plane $Al_2O_3$ had different level of impurities with a higher oxygen content and lower N/Sc ratio. Similar observations can be made for the film grown at 820 °C and at 700 °C, where the film grown on c-plane sapphire had the highest amount of oxygen impurities. This difference may originate from the type of epitaxial-like growth between the films. In one case (on c-plane sapphire), twin domain growth occurred but not in the other one (on MgO(111)). The presence of twin domain observed on ScN(111)/c-plane sapphire can be affiliated to defects/steps where adatom mobility (surface energy) is modified and favors the insertion of oxygen (here impurities). This phenomenon has been observed, for example, on a crystal of $MgAl_2O_4$ where impurities (Si, Ti) insertion arises at the twin boundaries of the monocrystal [49]. The difference in oxygen content in the two films oriented (001) and (111) grown on MgO may originate from the orientation of the film and the surface terminated during growth, where, in the case of ScN, a (001) terminated surface will be less favorable to oxygen (impurities) insertion compared to a (111) terminated surface. High percentage of oxygen lead to the formation of oxide into the film located most likely at the grain boundary and defects and may deteriorate the overall electrical properties of the films.

The electrical resistivity is decreased by a factor 10 when the $T_D$ is increased from 700 °C to 950 °C, and the absolute Seebeck coefficient values measured at room temperature are increased by 40%.

Different hypothesizes can be proposed to explain the improvement of the electrical properties with the temperature: 1) a densification of the film along with a morphology evolution; 2) an increase of the N/Sc ratio towards 1 along with a decrease of oxygen impurities (oxide/oxynitride), 3) the increase of the epitaxial orientation quality of the film.

At the highest $T_D$ (950 °C), differences between substrates is only noticeable on the composition of the film. Indeed, epitaxial orientation quality, morphology, densification, were comparable between samples. The three-films exhibited equivalent Seebeck coefficient (within the error bar) but different electrical resistivities. The film exhibiting a high electrical resistivity of about 200 µΩ cm was the one deposited on c-plane $Al_2O_3$. This electrical conductivity is equivalent to the values reported in the literature between 200 - 800 µΩ.cm (or $\sigma = 1\times10^3$ to $5\times10^3$ S cm$^{-1}$) [14-16, 21, 50]. The electrical resistivities of the two other films grown are reduced to values around 50 µΩ cm (or $\sigma = 2\times10^4$ S cm$^{-1}$) which is an improvement of 400%. The increase of the electrical conductivity of the films grown on MgO(111) and r-plane $Al_2O_3$ was related to the growth mode without twin domain on MgO(111) and r-plane sapphire leading to the absence of oxide in contrast to the films on c-plane $Al_2O_3$. Indeed, the film grown on c-plane sapphire was, at that temperature, the only one having a non-negligible amount of oxynitride and oxide located most likely at the grain boundary. Oxygen dissolved into ScN cell may act as donor and shift the Fermi level of ScN into the conduction band and increase the electrical conductivity, in contrary to the presence of oxide at the grain boundary which was proved to reduce the electrical conductivity [14, 20, 21, 30, 51, 52].

For the same $T_D$, Seebeck values are estimated around -40 µV K$^{-1}$ and comparable to values reported in the literature for the ScN films (from -25 to -60 µV K$^{-1}$ at 50°C) [14-16, 21, 50]. An estimation of the power factor ($S^2\sigma$) at room temperature was, at the best around 3.2 $\times10^{-3}$ W m$^{-1}$ K$^{-2}$ for the film grown at 950 °C on MgO(111) and r-plane $Al_2O_3$. To our knowledge, this value of power factor, at room temperature, is higher than the ones reported on previous works (at the best values from $1\times10^{-3}$ [14, 15] to $2\times10^{-3}$ W m$^{-1}$ K$^{-2}$ [21]). The key point of this improvement relies the control and reduction at the minimum amount of oxygen impurities incorporation (here 3-4%) to benefit from the doping effect expected by theory [17, 21, 30]. higher amount of oxygen into the films as oxynitride or oxide at the grain boundary inhibit the doping effect and deteriorate the overall electrical conductivity.

## Conclusion

In conclusion, epitaxial-like growth of ScN rock-salt structure were obtained on three different substrates with a (111) orientation on c-plane $Al_2O_3$ and MgO(111) and a (001) orientation on r-plane $Al_2O_3$. The increasing of the deposition temperature increased the crystal quality with a higher epitaxy of the film, smoother and denser film on all three substrates. XPS analysis allowed to detect the presence of oxynitride and oxide of scandium which was connected to the adatom mobility during the growth process affected by the temperature of deposition and the presence or not of twin domain growth induced by the orientation and nature of substrate higher oxygen contaminations was observed on ScN(111)/ /c-plane sapphire with the presence of twin domain in contrary to others films ScN(111)/MgO(111) and ScN(001)/r-plane sapphire.

At the highest deposition temperature of 950°C, where the films deposited on the three substrates differs by the amount of oxygen contaminants, a reduction the amount of oxygen impurities to 3 % decreased the electrical resistivity by a factor 4 while maintaining the Seebeck coefficient values. These samples exhibited low electrical resistivity for ScN films, with room-temperature values of 50 $\mu\Omega.cm$ and a Seebeck coefficient values of -40 $\mu V.K^{-1}$ which gives an estimated power factor of $3.2 \times 10^{-3}$ $W.m^{-1}K^{-2}$ at room temperature.

## Acknowledgments

The authors acknowledge the funding from the European Research Council under the European Community's Seventh Framework Programme (FP=2007–2013) ERC Grant Agreement No. 335383, the Swedish Foundation for Strategic Research (SSF) through the Future Research Leaders 5 program and the Swedish Research Council (VR) through project grant number 2016-03365, the Knut and Alice Wallenberg Foundation through the Wallenberg Academy Fellows program, and the Swedish Government Strategic Research Area in Materials Science on Functional Materials at Linköping University (Faculty Grant SFO-Mat-LiU No. 2009 00971). GG acknowledges the financial support from the Åforsk Foundation Grant 16-359, and the Carl Tryggers Stiftelse contract CTS 17:166.

## Conflicts of interest

There are no conflicts of interest to declare.

# Supplemental material

**XPS FITTING DETAILS**

Peak fitting was performed for all photoelectron peaks after a Shirley background subtraction and as described as follow using Voigt function for all components:

N 1s peak

1. The mathematical function is a Voigt function with a fixed shape defined in KolXPD by $FWHM_{gaussian}/FWHM_{lorentzian} = 1.15$. The full width that half maximum (FWHM) of identified peak component were kept fixed for all fitting: ScN (0.90 eV), $ScO_xN_y$ (1.5 eV). The energy separation between the different components were kept constant between the nitride and the oxynitride (+ 0.50 eV).

O 1s peak

1. The mathematical function is a Voigt function with a fixed shape for each component and defined by $FWHM_{gaussian}/FWHM_{lorentzian}$ equal to 1.20, 5.50 for the $ScO_x$ and $ScO_xN_y$, respectively. The full width at half maximum (FWHM) of the different components were kept fixed for all fitting: $ScO_x$ (1.10 eV) and $ScO_xN_y$ (2.03 eV). The contamination component was without any FWHM constraints with a FWHM varying between from 1.85 to 2.15 eV.
2. The energy separation between the oxynitride / oxide and the oxynitride / contaminant components (adsorbed species, gas) were kept constant at (1.60 eV and 3.20 eV, respectively) with a FWHM varying between 1.85 eV and 2.05 eV for the contaminant peak component.

Sc 2p peak

1. The spin-orbital splitting of Sc 2p peak was respected for all different components with intensity ratio fixed between the $2p_{3/2}$ and $2p_{1/2}$ peaks and the relative distance between the Sc $2p_{3/2}$ and Sc $2p_{1/2}$ peaks fixed at 4.50 eV.
2. The mathematical function is a Voigt function with a fixed shape for the ($ScN_x$), the nitride (ScN) and the oxide ($ScO_x$) component and defined by $FWHM_{gaussian}/FWHM_{lorentzian}$ equal to 2.35. The full width at half maximum (FWHM) of the oxide component was kept fixed for all fitting: $ScN_x$ (0.88 eV), ScN (1.58 eV) and $ScO_x$ (1.60 eV)
3. The energy separation between the different components were kept constant at 0.98 eV ($ScN_x$-ScN), at 2.50 eV (ScN-$ScO_x$).
4. Due to possible level of oxygen into the oxynitride, the corresponding component was deconvoluted from the other components as peak with less constraint. The energy separation between the nitride and oxynitride component varies from 1.45 to 1.98 eV when the oxygen content in the film increases and the FWHM varies from 1.30 to 1.62 eV.

**EPITAXIAL RELATIONSHIP**

Fig. S1 presents the epitaxial relationship between ScN and the different substrates: c-plane $Al_2O_3$, MgO(111) and r-plane $Al_2O_3$. The films were (111) epitaxially grown on c-plane $Al_2O_3$ and MgO(111) and (001) epitaxially grown on r-plane $Al_2O_3$.

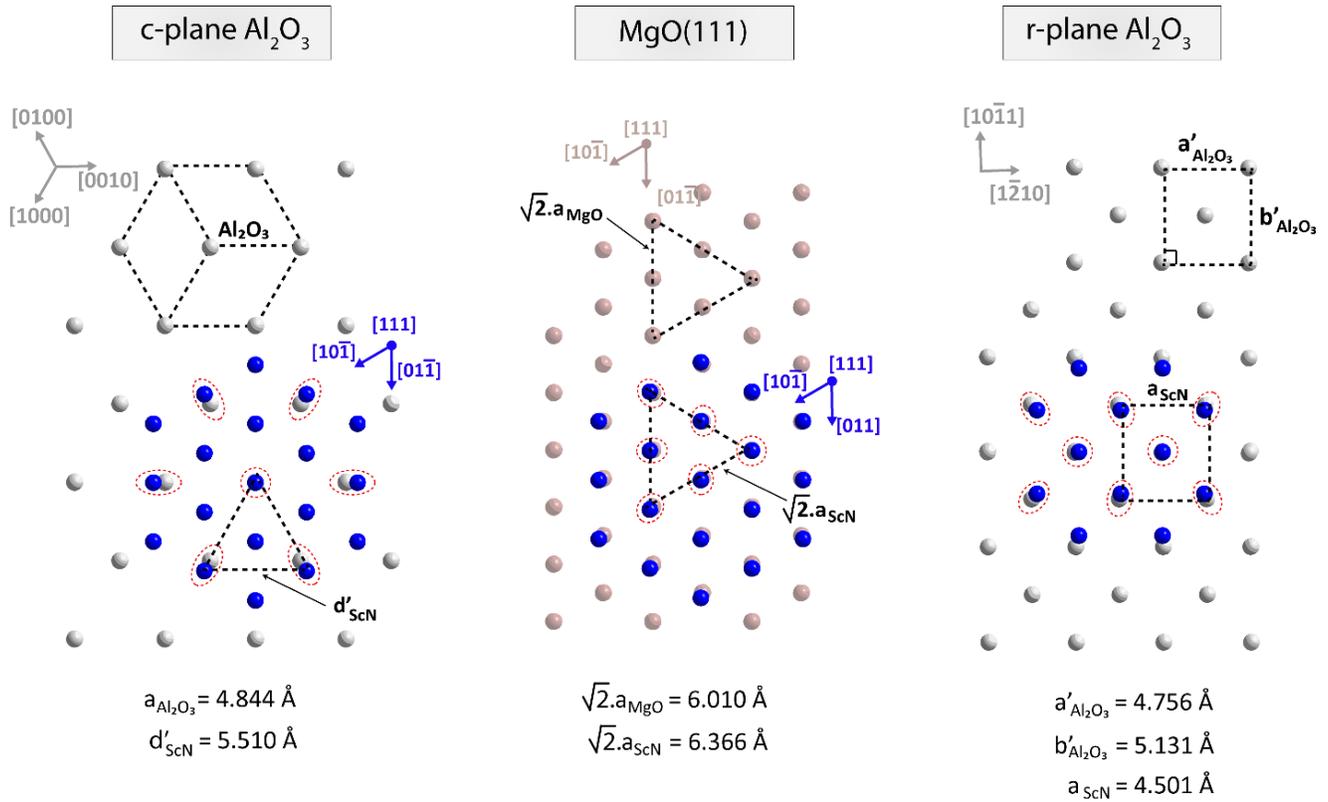

**Fig. S1:** Schematic view of the epitaxial relationship between ScN and c-plane $Al_2O_3$, MgO(111) and r-plane $Al_2O_3$. The ScN is represented by the scandium ions (blue); The $Al_2O_3$ is represented by the aluminum ions (grey); and the MgO is represented by the magnesium ions (light brown).

**XPS C 1s CORE LEVEL SPECTRA**

Fig. S2 presents the C 1s core level spectra after sputter cleaning of the nine ScN films deposited at three different temperatures on three different substrates. The nine films were subject to a same sputter cleaning process before analysis. The carbon detected at the surface of the film after sputter cleaning differs between substrates and temperatures of deposition. The effectiveness of the cleaning in this study was attributed mostly to the surface morphology varying between the different films with a higher effectiveness when the film had a smother surface.

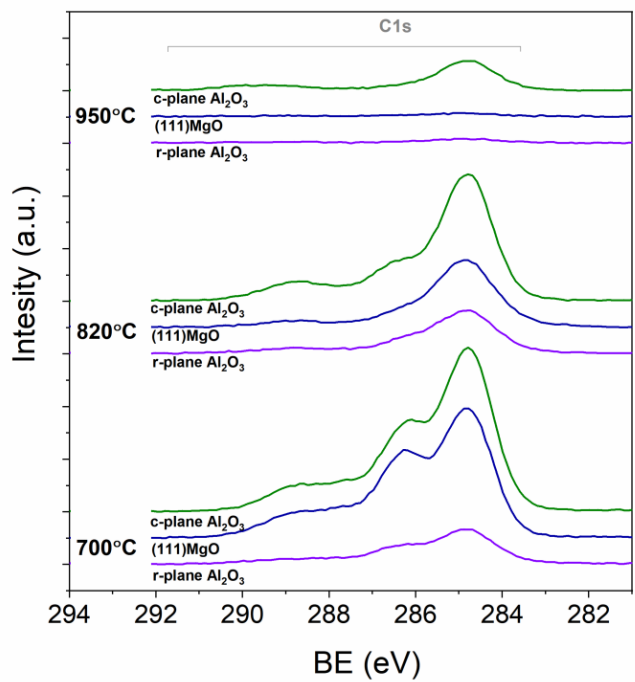

**Fig. S2:** XPS measurement of the C 1s core level, after sputter cleaning process, measured on films deposited on different substrates (c-plane $Al_2O_3$, MgO(111) and r-plane $Al_2O_3$) and at different temperatures (700°C, 820°C and 950°C).